\documentclass[prb,twocolumn,amsmath,amssymb]{revtex4}

\usepackage{graphicx}
\usepackage{dcolumn}
\usepackage{bm}
\usepackage{epstopdf}

\def\be{\begin{equation}}
\def\ee{\end{equation}}
\def\bd{\begin{displaymath}}
\def\ed{\end{displaymath}}
\def\-{\phantom{-}}

\begin{document}

\title{Voltage-dependent spin flip in magnetically-substituted graphene nanoribbons: Toward the realization of graphene-based spintronic devices}

\author{Gregory Houchins$^1$}
\author{Charles B. Crook$^1$}
\author{Jian-Xin Zhu$^{2,3}$}
\author{Alexander V. Balatsky$^{4,5}$}
\author{Jason T. Haraldsen$^{1,6}$}
\affiliation{$^1$Department of Physics and Astronomy, James Madison University, Harrisonburg, VA 22802}
\affiliation{$^2$Theoretical Division, Los Alamos National Laboratory, Los Alamos, NM 87545, USA}
\affiliation{$^3$Center for Integrated Nanotechnologies, Los Alamos National Laboratory, Los Alamos, NM 87545, USA}
\affiliation{$^4$Institute for Materials Science, Los Alamos National Laboratory, Los Alamos, NM 87545, USA}
\affiliation{$^5$NORDITA, Roslagstullsbacken 23, 106 91 Stockholm, Sweden}
\affiliation{$^6$Department of Physics, University of North Florida, Jacksonville, FL 32224}
\date{\today}

\begin{abstract}

We examine the possibility of using graphene nanoribbons (GNRs) with directly substituted chromium atoms as spintronic device. Using density functional theory, we simulate a voltage bias across a constructed GNR in a device setup, where a magnetic dimer has been substituted into the lattice. Through this first principles approach, we calculate the electronic and magnetic properties as a function of Hubbard U, voltage, and magnetic configuration. By calculating of the total energy of each magnetic configuration, we determine that initial antiferromagnetic ground state flips to a ferromagnetic state with applied bias. Mapping this transition point to the calculated conductance for the system reveals that there is a distinct change in conductance through the GNR, which indicates the possibility of a spin valve. We also show that this corresponds to a distinct change in the induced magnetization within the graphene.

~

\noindent Corresponding Author: Dr. Jason T. Haraldsen (j.t.haraldsen@unf.edu)


\end{abstract}
\maketitle

\section*{Introduction}

Over the last couple of decades, solid-state research has guided the advancement of technologies, which includes the control and manipulation of charge transport in various materials. Progress in the overall understanding of electronic systems has led to the development of p-n junctions, giant and colossal magnetoresistance, and the whole realm of semiconductor physics, for which today's technology is based. Currently, the vast majority of technologies take advantage of only the charge degree of freedom in materials. However, over the last decade, there has been a large push to combine the electronic and spin degrees of freedom in both multiferroic materials and spintronic devices \cite{wolf:01,bade:10,zuti:04}. These endeavors have ranged from molecular spintronics \cite{roch:05,boga:08,khaj:11} to quantum dot systems \cite{enge:01} to topological insulators \cite{pesi:12}, where a major motivation is the realization and enhancement of quantum computation and nanotechnology \cite{awsc:02,mats:03,xu:06}.

Recently, graphene has gained a large amount of experimental and theoretical attention due to its distinct electron mobility produced produced thorugh carbon-carbon (C-C) $\pi$ bonding that enables an electron delocalization throughout its honeycomb lattice structure\cite{geim:07,neto:09,chen:08,pop:12,bolo:08}. Due to the electronic structure within graphene, known as a Dirac material, where the presence of a Dirac cone in the electronic structure along the $K$ direction\cite{wehl:14} allows graphene to be classified as both a zero-gap semiconductor and a zero density of states metal\cite{geim:07,novo:05}. This means that the electrons have the characteristics of ultra-relativistic massless particles\cite{mull:09}. Furthermore, graphene also exhibits enhanced thermal and tensile strength properties\cite{moro:08,bala:08,zhu:10}.

\begin{figure}
\includegraphics[width=1 \linewidth]{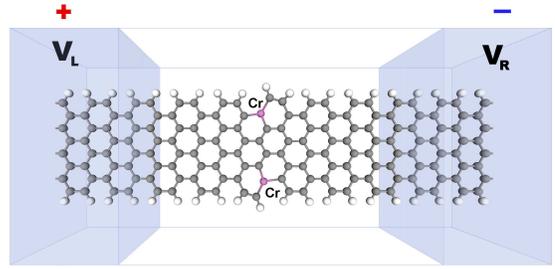}
\caption{A GNR device setup between two electrodes. The voltage between the two electrodes is varied by giving the left electrode a positive voltage and the right a negative voltage each equal in magnitude to half the total potential difference desired. The graphene is then doped with two chromium atoms separated by four carbons.}
\label{setup}
\end{figure}

Besides the well-known electronic and thermal properties of graphene, it has been shown that graphene can obtain distinct magnetic properties through the addition of adatoms placed on the lattice or through the placement of graphene on YIG and various other interactions\cite{yazy:08,rao:12,sant:10,hu:11,wang:15,wang:12,lv:12,duga:06}. Recently, it was showed that the direct substitution of magnetic impurities into the graphene lattice can also induce localized magnetism that interacts through the conduction electrons in the graphene\cite{Boukhvalov:08,crook:15}. Furthermore, this work, along with others, have detailed the potential for using graphene substitution as spintronic transistors, spin valves, and other devices\cite{Zhang:14,Sun:13}. 

In our previous computational work\cite{crook:15}, the presence of two transition-metal atoms in graphene produced a distinct induction of magnetism in local carbon atoms, where it was shown that the spatial effect of two magnetic atoms produced RKKY interactions through the conduction electrons that culminated in either ferromagnetic or antiferromagnetic ground states\cite{crook:15,sher:11}. Here, the ability to induce magnetism and have a RKKY-dependent interaction signals to the potential for transition-metal substituted graphene nanoribbons (GNRs) as a possible spintronic device. While the use of GNRs as a spintronic device has been demonstrated in various studies\cite{Zhang:14,Sun:13,Haney:09}, these studies have mainly focused on the utilization of magnetic edge states in the GNR. Therefore, it is proposed that direct substitution of the magnetic atoms may provide better control and handling of the spin states through an applied voltage coupling to the RKKY interactions.

\begin{figure}
	\includegraphics[width=0.9 \linewidth]{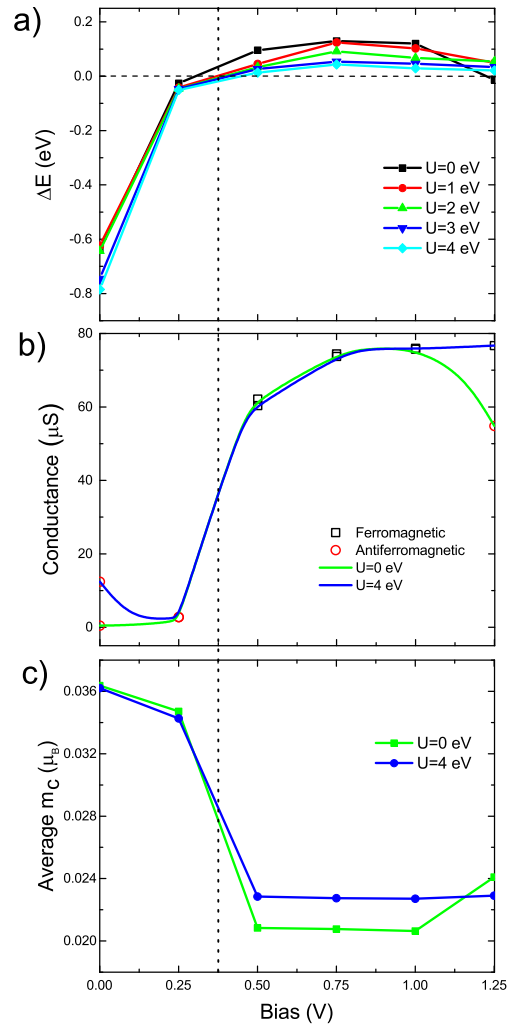}
	\caption{(a) The difference in total energy of the two Cr spin orientation states. Here, the difference is taken to be up-down minus up-up therefore a positive difference denotes a spin up-up ground state. This difference is shown for various Hubbard U varying from 0 eV to a max of 4 eV. (b) The calculated conductance of the GNR device as a function of bias for both U= 0 eV and U= 4 eV. (c) This shows the average of the absolute value of the magnetic moment of all the carbon atoms as a function of bias for a Hubbard U of 0 eV and 4 eV. There is a vertical dotted line at 0.375 V to show the same point of change in characteristics from AFM to FM in all three graphs.}
	\label{exchange}
\end{figure}

In this study, we explore the possibility of controlling the magnetic states generated by two chromium atoms substituted into graphene through the use of an applied bias voltage. Using density functional theory through a generalized gradient approximation, we simulated a spintronic device made from a GNR with two chromium atoms substituted into the lattice. By calculating multiple magnetic configurations using varying on-site potential and applied voltage bias, we determined the device density of states, magnetic profile, and voltage dependence for these spintronic devices. Furthermore, we examined the conductivity of the GNR and show that as the nanoribbon crosses a critical voltage, the magnetic ground state flips from antiferromagnetic to ferromagnetic, which produces a distinct change in the induced magnetism of graphene and the overall conductance. This provides a theoretical realization of a graphene-based spintronic device and will hopefully motivate experimental endeavors in this direction. 

\begin{figure}
	\includegraphics[width=1 \linewidth]{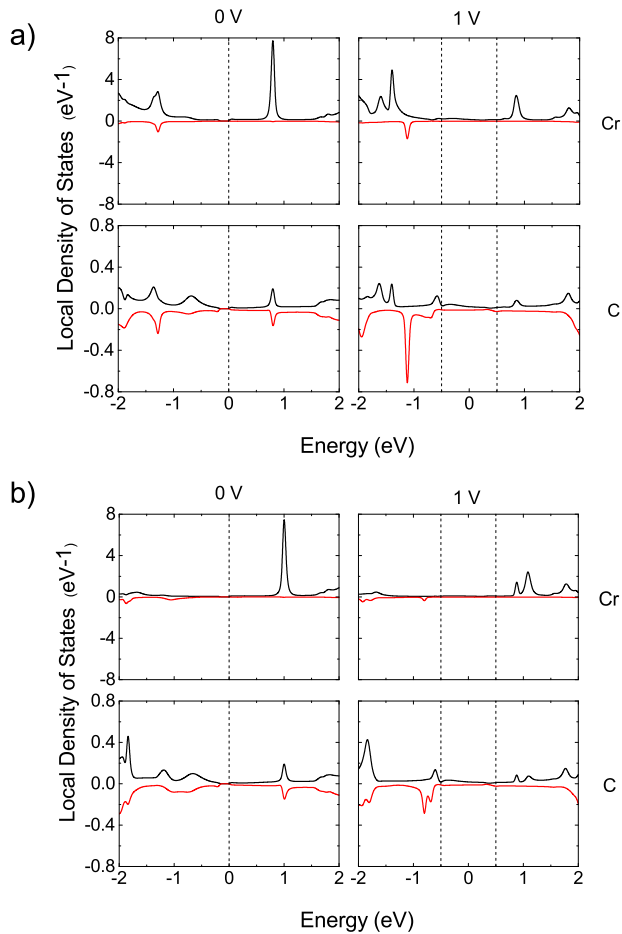}
	\caption{The local density of states for Chromium and Carbon with Hubbard U of (a) U= 0 eV and (b) U= 4 eV for a bias of 0 and 1 V each. The black line denotes the spin up electron states while the red line denotes spin down. The dotted lines denote the Fermi levels of the device. Note in the case of a 1 V bias, there are two Fermi levels at -0.5 eV and 0.5 eV due to the bias. The atoms used for this figure are the top Cr atom and the second carbon atom down from that Cr with respect the orientation seen in FIG. \ref{setup}.}
	\label{local-dos}
\end{figure}

\begin{figure}
	\includegraphics[width=1 \linewidth]{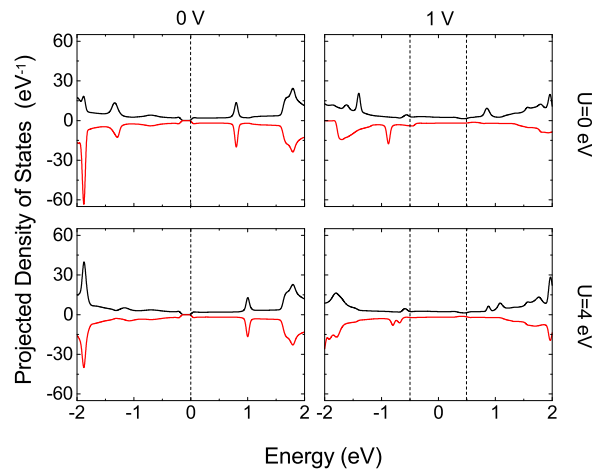}
	\caption{The projected density of states for a bias of 0 and 1 V both in the case of a Hubbard U of 0 eV and 4 eV. Again the black denotes spin up and red denotes spin down. The Fermi levels are again shown with dotted lines.}
	\label{projected-dos}
\end{figure}

\section*{Antiferromagnetic/Ferromagnetic Spin Flip}

In a previous study, it was shown that the magnetic interactions between chromium atoms in graphene interact through an induction of magnetism in graphene and couple through the conduction electrons in the carbon\cite{crook:15}. This indicated that there may be the potential to affected the magnetic states through the use of an external electric field or applied voltage. Therefore, this study presents the results from a device setup to show that two Cr atoms in a GNR may have the ability to produce spin switch, which opens the possibility for creating a spintronic device using magnetically-doped graphene.

To create a spin interaction, two chromium atoms are placed in the GNR along the zig-zag direction with a spacing of four carbon atoms. $\Delta E$ is determined by comparing the calculated total energy for the antiferromagnetic ($\uparrow\downarrow$) and ferromagnetic ($\uparrow\uparrow$) Cr-Cr configurations. Considering a classical dimer with a Hamiltonian given by

\be
H = -\frac{J}{2} \bar S_1\cdot \bar S_2,
\ee

where $J$ is the superexchange energy ($J > 0$ for FM and $J < 0$ for AFM) and $S$ is the spin\cite{hara:05}. The classical energy is

\be
E = -\frac{JS}{2}^2cos(\theta_1-\theta_2),
\ee
where $\theta$ = 0$^{\circ}$ or 180$^{\circ}$ for up or down spins, respectively. A general $\theta$ can be used for frustrated or canted spins\cite{hara:09}. From this, $E_{FM} = -JS^2/2$ and  $E_{AFM} = JS^2/2$, which makes $\Delta E = JS^2$. Therefore, the change in energy is directly related to the Cr-Cr dimer superexchange energy. It should be noted that this is a generalization, since the induction of magnetic moments in the near by carbon atoms will produce a magnetic ''dumb-bell". However, the size of the induced moment is small and will only provide a small perturbation of the original Hamiltonian. 

As shown in Fig. \ref{exchange}(a), the zero-bias magnetic ground state for the device setup is determined to be antiferromagnetic, which is consistent with the bulk supercell calculations from Ref. [\onlinecite{crook:15}]. Furthermore, as the applied voltage bias across the device is increased, there is a distinct change in the magnetic ground state, at $V$ = 0.37 V, where the ground state is shifted to a ferromagnetic state. 

At the same critical voltage, we find that conductance shows a dramatic increase throughout the GNR(Fig. \ref{exchange}(b)), while producing an overall drop in the average magnetic moment in carbon (Fig. \ref{exchange}(c)). This is consistent regardless of the assumed onsite Hubbard U. As the voltage further increases, the simulation of the $U = 0$ eV device oscillates back to an antiferromagnetic ground state. This could be indicative of the interaction of the conduction electrons and the voltage bias, which produces a voltage dependent interaction as observed in Ref. [\cite{Haney:09}]. However, as the on-site potential is increased, the overall oscillatory nature is reduce and practically levels out. This provides a potential experimental verification point. 
 
For the $U$ = 0 calculations, the conductance drops again around 1.25 eV, which indicates that this phenomena may be dependent on the AFM state. The low conductance in the AFM state below the critical voltage is consistent with a magnetoresistance device. Therefore, the use of conductance could be used to determine that a spin flip has occurred. From these calculations, there is a threshold of about 60 $\mu$S indicating a transition from AFM to FM.

\begin{figure}
	\includegraphics[width=1 \linewidth]{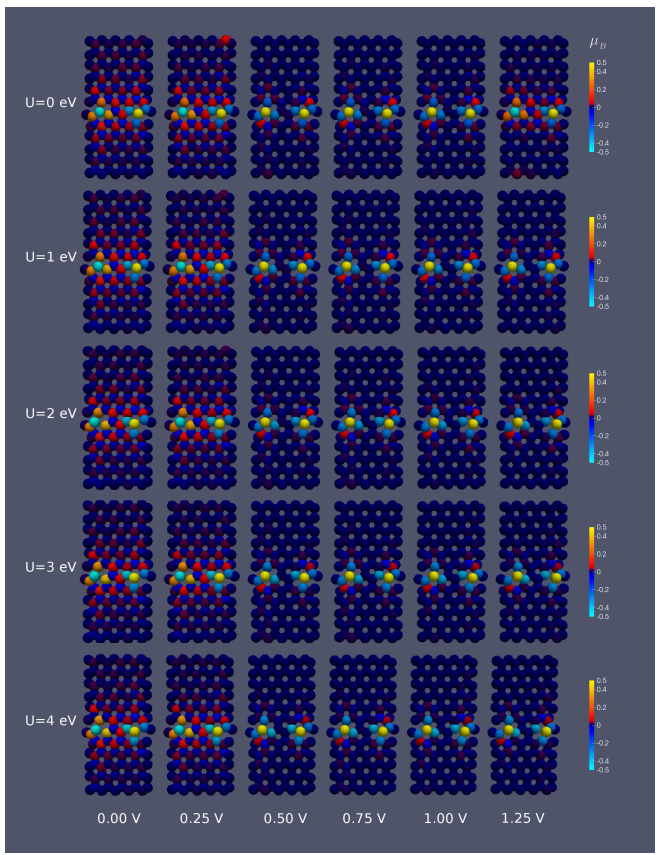}
	\caption {Magnetic moment for each atom in the GNR device for various biases and Hubbard U. The scale has been renormalized between -0.5 $\mu_{B}$ and 0.5 $\mu_{B}$ to show the magnetization of the carbon atoms around the chromium in more detail. The magnetic moment on the Cr varies from 3.6 to 4.0 $\mu_{B}$ depending on the Hubbard U as seen in Fig. \ref{cr-moment}.}
\label{magnetization}
\end{figure}

\section*{The Device Density of States}

The shift in magnetic state indicates that, as the voltage is increased, the electrons used to produce the antiferromagnetic interaction are elevated into the conduction band. This is shown in the calculated density of states (DOS) (shown in Fig. \ref{local-dos}), where the change in the electronic structure shifts the magnetic interaction between the chromium atoms and produces a ferromagnetic ground state configuration. Figure \ref{local-dos}(a) shows the calculated local DOS for both the Cr and C sites at both 0 and 1 V for a Hubbard U of 0 eV (Fig. \ref{local-dos}(b) shows U = 4.0 eV). A comparison of the projected DOS for both U = 0 and 4 eV (shown in Fig. \ref{projected-dos}) shows that this effect is produced regardless of an on-site potential.The density of states of the device under no bias, there are clearly no states at the Fermi level and therefore no conduction. This changes as you increase the on-site potential. However, at a bias of 1.0 V, the number of states are increase at the Fermi level.

\begin{figure}
	\includegraphics[width=1 \linewidth]{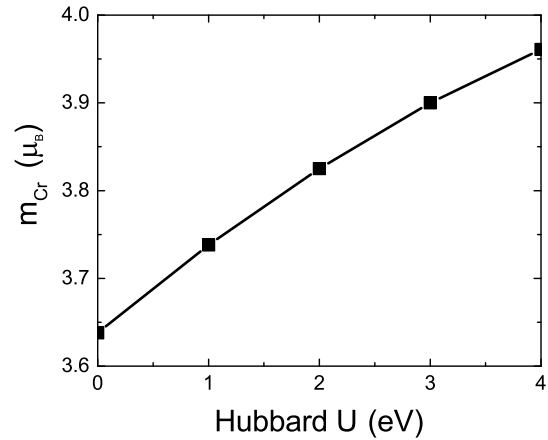}
	\caption {The magnetic moment for the top Chromium atom in the orientation of Fig. \ref{setup} as a function of Hubbard U. For simplicity the magnetic moment of Cr in the case of no bias was used since the magnetic moment varied little with respect to bias.}
	\label{cr-moment}
\end{figure}

If we look at the local density of states for just one chromium atom, we see a similar pattern of no states at the Fermi level without a bias present and then states become present when the bias is added. This means that the conduction of the graphene increases as the voltage increases, which is supported by the calculated conductance of the device at each bias. There is a direct correlation of conductance and magnetic ordering as seen in Fig. \ref{exchange} indicating that the increased conductance is producing the flip in exchange interaction between the chromium atoms.  We can see similar patterns in the local density of states for the carbon, however the number of states are about an order of magnitude less than that of Cr.

\section*{Magnetization Mapping}

In Fig. \ref{magnetization}, we determine the magnetic moment as function of atomic spacing throughout the simulated GNR along with varying voltage and on-site potential. The first two columns show a consistent AFM configuration for the Cr-Cr dimers. However, it is clear that a distinct magnetization is drawn into the nearest neighbor carbon atoms, which was shown a proximity induced magnetization of graphene in Ref. [\onlinecite{crook:15}]. This induction of magnetic moment on the carbon atoms is reduced slightly with increasing on-site potential, which shown as an increase in the Cr magnetic moment as a function of $U$ in Fig. \ref{cr-moment}. As applied voltage is increased through the critical voltage of 0.37 V, there is a dramatic shift in the magnetization profile as conductance is increases and the Cr-Cr dimer transitions to a FM state. There is also a shift in the magnetic moment of the carbon (shown in Fig. \ref{exchange}(c)).

The decrease in magnetization in the FM states implies that the magnetization is inversely related to conductance, which the biggest decrease in magnetism in the two middle carbons between the Cr. This seems to be produced by the shift in conductance which is affect by the voltage-dependent magnetic interaction and supports the conclusion that conduction electrons are mediated through a RKKY interaction\cite{rude:54,kasu:56,yosi:57}.

\section*{Conclusions}

Overall, we examine the voltage-dependence of magnetically-substituted GNR device, and show that the magnetic interactions change with applied voltage strength. Based on an analysis of the projected DOS and the conductivity, it appears that the coupling interaction between the Cr is competing with the conduction electrons that contribute to antiferromagnetic ordering. Once the critical voltage is reached, the magnetic coupling between the Cr atoms flips into an overall FM interaction. This changes that magnetic induction within the GNR; changing it from AFM to FM as well. As shown in Ref. [\onlinecite{crook:15}], the FM state allows for a greater conduction of the carbon atoms (shown by the LDOS in Fig. \ref{local-dos}). Therefore, the overall conductance is increase by a few orders of magnitude.

The dramatic change in the conductance of the device at the same critical voltage for which there is a change in the magnetic state of the chromium atoms allows for a distinct observable and the potential realization of graphene-based spintronic device. This correlation is supported by the apparent decrease in conductance when the U = 0 eV case flipped back to AFM ground state. Essentially, the AFM ground state forces the carbon atoms into an induced AFM state, which changes the reduces to the conductance by producing a correlation barrier. Once the GNR is changed into the FM state, the correlation barrier is reduced or eliminated.

In closing, this study shows the potential for GNRs to be utilized as potential spintronic devices with direct magnetic substitutions. It is the goal of this research to motivation experimental endeavors in this direction.

\section*{Acknowledgements}

This work was supported by U.S. DOE Basic Energy Sciences E304. G.H., C.B.C., and J.T.H. acknowledge undergraduate support from the Institute for Materials Science at Los Alamos National Laboratory. The work at Los Alamos National Laboratory was carried out under the auspice of the U.S. DOE and NNSA under Contract No. DEAC52-06NA25396 and supported by U.S. DOE Basic Energy Sciences Office (J.-X.Z and A.V.B).

\section*{Author Contributions}
J.T.H., J.-X.Z., and A.V.B conceived the research. G.H., C.B.C., and J.T.H. carried out the numerical simulations. J.T.H., G.H., C.B.C., J.-X.Z., and A.V.B. discussed the results and co-wrote the paper.

\section*{Competing financial interests}

The authors declare no competing financial interests.

\section*{Computational Methods}

This report presents first principles calculations of electron interactions throughout a magnetically-doped GNR device.  To simulate a spin switch device, we constructed a GNR (10.4 \AA by 22.7 \AA) with  0.710 nm electrodes on each side consisting of 120 total atoms. The electrodes are used to simulate a voltage bias across the device. In the center of the GNR, two of the carbons were replaced by chromium atoms, where the chromium atoms were separated by 4 carbon atoms along the zig-zag chain direction. To assure a nanoribbon configuration, the dangling bonds of the edge carbons are capped with hydrogen. This is illustrated in Fig. \ref{setup}. 

Using the density functional codes of Atomistic Toolkit\cite{quantumwise}, we performed a geometry optimization using a 3x3x3 $k$-point optimization. This allowed us to obtain the known graphene impurity distortion shown in Crook et al.\cite{crook:15}. Once the geometry was precisely determined, this was kept constant due to the computational expense of the geometric optimization for the device setup. The electronic structure was then optimized using a spin-polarized generalized gradient approximation (SGGA) with Perdew, Burke, and Ernzerhof functionals and 5x5x5 $k$-point mesh\cite{bran:02,sole:02,perd:96}. Calculations were performed for various onsite potentials U (0, 1, 2, 3, and 4 eV) and voltage biases (0, 0.25, 0.5, 0.75, 1 V). The overall magnetic ground state for each set of parameters was determined by comparing the total energy for a ferromagnetic and antiferromagnetic configuration. To assure consistency between magnetic ground states, we used a tolerance of 0.3 meV for the total energy convergence. Furthermore, to examine the electronic and magnetic properties, we calculated the device density of states, Millikan population (magnetic moment), and conductance.

\end{document}